\documentclass[conference]{IEEEtran}
\IEEEoverridecommandlockouts

\usepackage{cite}
\usepackage{amsmath,amssymb,amsfonts}
\usepackage{algorithmic}
\usepackage{graphicx}
\usepackage{subcaption}
\usepackage{booktabs}
\usepackage{textcomp}
\usepackage{xcolor}
\usepackage[colorlinks=true,allcolors=blue]{hyperref}
\usepackage{url}
\usepackage{listings}

\lstdefinestyle{pystyle}{%
  language=Python,
  basicstyle=\ttfamily\footnotesize,
  keywordstyle=\color{blue!70!black},
  commentstyle=\color{gray},
  stringstyle=\color{green!45!black},
  showstringspaces=false,
  breaklines=true,
  columns=fullflexible,
  frame=single,
  framesep=4pt,
  aboveskip=4pt,
  belowskip=2pt,
}

\graphicspath{{figures/}}

\newcommand{\dcx}{\Delta_{\mathrm{CX}}}
\newcommand{\cxm}{\mathrm{CX}_{\mathrm{match}}}
\newcommand{\cxp}{\mathrm{CX}_{\mathrm{Pauli}}}
\newcommand{\ctqw}{CTQW}
\newcommand{\pms}[2]{$#1\,{\scriptstyle\pm#2}$}

\begin{document}

\title{Predicting Resource Efficient Hamiltonian Decomposition for Continuous-Time
Quantum Walk Simulations}

\author{%
\IEEEauthorblockN{Mostafa Atallah\IEEEauthorrefmark{1},
Rebekah Herrman\IEEEauthorrefmark{1}, and
Zain H.\ Saleem\IEEEauthorrefmark{2}
}
\IEEEauthorblockA{%
\IEEEauthorrefmark{1}University of Tennessee, Knoxville\\
\IEEEauthorrefmark{2}Argonne National Laboratory\\
Email: matalla3@vols.utk.edu}%
}

\maketitle

\begin{abstract}
Simulating a continuous-time quantum walk (\ctqw) on a graph in the circuit model of
quantum computing requires decomposing its Hamiltonian into terms that can be
Trotterized into hardware-native gates. We consider two such decompositions: the
standard Pauli decomposition and the recently introduced matching decomposition. Recent work suggests that matching decomposition requires fewer CX gates than Pauli decomposition when simulating CTQWs on sparse graphs, while Pauli decomposition requires fewer CX gates on denser graphs. 
Since CX gates dominate error and runtime on current hardware, we train machine learning models to predict, for a given
graph on which we want to perform a CTQW, which of the two decompositions produces
the smaller CX gate count. We train and evaluate on the complete population of
all $11{,}117$ connected eight-vertex graphs from Brendan McKay's simple graph database. Because this is the entire population rather than a sample, the class balance and the overlap between the two classes are measured directly
rather than estimated from a subset. We train on twelve features, ten topological
properties of the graph and two that count the number of terms the Pauli and matching decompositions produce
($n_{\mathrm{Pauli}}$ and $n_{\mathrm{match}}$). These two quantities can be calculated without fully transpiling the CTQW simulation circuit in the gate model. Surprisingly, standard topological properties of the graph alone provide little predictive power. Instead, the dominant predictive signal comes from $n_{\mathrm{Pauli}}$, which is a property of the Hamiltonian decomposition rather than an intrinsic property of the graph.  Degree variance (the
statistical variance of the vertex-degree sequence) is the only other feature
that carries signal, and the remaining ten features are near zero. %The two matter for
%different reasons: $n_{\mathrm{Pauli}}$ alone reaches a Matthews correlation
%coefficient (MCC) of $0.423$, and adding degree variance raises it to $0.574$,
%close to the $0.593$ of the full twelve-feature set.
We test a range of machine
learning models on the dataset. Their Matthews correlation coefficients (MCC) fall in a
narrow band, from $0.569$ for the best untuned model to $0.593$ after tuning, so no
single architecture stands out. We adopt a single-hidden-layer artificial neural
network (ANN) with $32$ nodes as the selected model, at MCC $0.593$, where every deeper
network overfits the $418$ minority graphs. Applied
frozen to a held-out, class-balanced test set of graphs on $N\in\{2^n\}_{n=3}^8$
vertices drawn from structured and Erd\H{o}s--R\'enyi families, the tuned model
transfers, with MCC rising from $0.785$ at $N=8$ to $1$ at $N\ge64$.
\end{abstract}


%==============================================================================
\section{Introduction}
%==============================================================================
Continuous-time quantum walks (CTQWs) are a fundamental primitive for quantum
algorithms in search, sampling, and graph
problems~\cite{kempe2003quantum,venegas2012quantum,childs2004spatial}, and
constitute a model of universal quantum
computation~\cite{childs2009universal}. A CTQW on a graph $G=(V,E)$ evolves under a
Hamiltonian $H$ derived from $G$, and throughout this work $H$ is the
\emph{adjacency matrix} of $G$. Furthermore, we enforce that each graph has $2^n$ vertices for $n \in \mathbb{N}$ and each vertex of $G$ is labeled by an $n$-digit bitstring. Running a CTQW on gate-based hardware
requires implementing the time evolution $e^{-iHt}$. For
a small class of highly structured graphs called circulant graphs this can be done exactly, as when $H$ is
circulant, $e^{-iHt}$ is diagonalized by the quantum Fourier
transform~\cite{qiang2016efficient,loke2017efficient}. However, general graphs admit no such
exact construction, and the standard approach is to \emph{decompose} $H$ into a sum
of terms that can each be exponentiated by a short gate sequence and to approximate
$e^{-iHt}$ by Trotterization over those
terms~\cite{childs2021theory,Suzuki1976_GenTrottFormAndSystApprox}. On near-term
hardware the cost of the resulting circuit is dominated by its two-qubit (CX) gate
count, which is set by the number and structure of the terms, so the way in which
$H$ is decomposed is the primary determinant of cost.

We compare two such decompositions, the standard Pauli decomposition and the
recently proposed matching decomposition~\cite{atallah2026simulating}, whose CX gate counts depend on different properties of the underlying graph. The \emph{Pauli decomposition} writes
$H=\sum_k c_k P_k$ as a sum of Pauli products, where each $P_k$ is a tensor product
of single-qubit Pauli operators,
\begin{equation}
P_k = \sigma_1^{(k)} \otimes \sigma_2^{(k)} \otimes \cdots \otimes \sigma_n^{(k)},
\qquad \sigma_j^{(k)} \in \{I, X, Y, Z\},
\end{equation}
and uses these products as the basis for
Trotterization~\cite{lloyd1996universal,whitfield2011simulation,childs2021theory}, so
the term count depends on the algebraic form of $H$. The number of Pauli terms is
only an approximate proxy for the CX gate count, however. A Hamiltonian of the form
$X_1 + X_2 + X_3 + \cdots + X_n$, for instance, has $n$ Pauli terms yet requires no
CX gates, since each term is a single-qubit rotation. The \emph{matching
decomposition}~\cite{atallah2026simulating} partitions the edges of $G$ into
matchings such that each edge in the matching has equal Hamming distance, defined as the number of digits that differ in the labels of the two vertices at the endpoint of each edge. Each matching is implemented as a block of
multi-controlled $R_x$ gates. Thus, the number of CX gates is governed by the graph's matching
structure~\cite{lovasz2009matching}. Throughout, we call a decomposition
\emph{cheaper} for a given graph when it produces the smaller CX gate count for
that graph. Because the Pauli cost depends on the algebraic form of $H$ while the
matching cost depends on the Hamming distance of each edge in the graph,
neither decomposition is cheaper for every graph. When the edges cluster at a few
Hamming distances the matching decomposition compresses them into few blocks and
produces fewer CX gates, whereas when the Hamming distances are spread out, as in
dense or random graphs, the Pauli decomposition produces fewer CX gates.

Determining which decomposition is cheaper exactly would require synthesizing and
transpiling \emph{both} circuits and comparing their CX counts,which can take a prohibitive amount of time for large graphs. We therefore want to learn a supervised
classification problem: given a graph, predict from inexpensive features which of
the two decompositions yields the smaller CX count, so that a compiler can commit to
it without building the circuit.

This problem sits between two established lines of research. Work on Hamiltonian
simulation reduces the cost of a decomposition that has \emph{already} been fixed,
through term grouping~\cite{whitfield2011simulation,gui2020term}, Trotter
ordering~\cite{tranter2019ordering}, minimum-clique-cover
grouping~\cite{verteletskyi2020measurement,crawford2021efficient}, randomization as in
qDRIFT~\cite{campbell2019random}, product-formula
design~\cite{childs2018toward}, and error bounds~\cite{childs2021theory}. Work
on circuit compilation reduces the cost of a circuit that \emph{already exists},
through qubit mapping and layout~\cite{maslov2008quantum,li2019tackling,tan2020optimal},
noise-adaptive mapping~\cite{murali2019noise}, and gate-count and non-Clifford
reduction~\cite{childs2017automated,kissinger2020reducing,sivarajah2021t}, increasingly
with the help of machine learning ~\cite{kusyk2021survey, fosel2021quantum,moflic2023towards, kremer2024practical, acampora2021deep,pozzi2022using,fan2022optimizing,paler2023machine, quetschlich2023predicting}. Both lines
take the decomposition as given.  In this work, we choose between two decompositions before either circuit is built, as this decision can have downstream optimization implications. To our knowledge it has not
previously been posed as a learning problem for CTQW simulation.

This paper makes four contributions. First, using IBM Qiskit at optimization
level~$3$, we compile CTQWs on all $11{,}117$ connected eight-vertex graphs under
both decompositions, recording the CX gate count and circuit depth for each, and
release the resulting labeled dataset. As this is the complete population of
eight-vertex graphs rather than a sample, the class balance and the overlap between
the two classes are determined exactly. Second, we and rank ten topological graph properties and the two term counts produced by the
decompositions by permutation feature importance and run a
feature-subset ablation in which the model is retrained on increasingly large
subsets of the twelve features, starting from the topological features alone and
adding the decomposition counts one at a time, under a common protocol so that only
the feature set varies. This separates the features that carry the predictive
signal from those that are redundant. Fourth, we tune a selector on the eight-vertex population
and evaluate it, without retraining, on a class-balanced held-out test set of graphs
of up to $256$ vertices, measuring how well the decision transfers beyond the
training size.

Two findings emerge. Among the twelve features, only the Pauli term count and,
secondarily, the variance of the vertex degrees carry predictive signal, so the
decision is governed by a quantity the decomposition already exposes rather than by
the graph's topology. The selector also transfers beyond its training size. While the selector was trained
on eight-vertex graphs, it predicts the more CX efficient decomposition on graphs an order of
magnitude larger with increasing accuracy, reaching perfect agreement at $N\ge64$. A
compiler can therefore choose the decomposition that can be implemented using fewer CX gates from a single
feature that is easy to calculate, avoiding the construction of the more costly circuit entirely.

The remainder of the paper is organized as follows. Section~\ref{sec:problem} states
the selection problem and the cost model. Section~\ref{sec:dataset} describes the
graph datasets, the labeling procedure, and the feature set. Section~\ref{sec:eda}
presents the feature-importance and class-overlap analyses. Section~\ref{sec:model}
describes the models and evaluation protocol, and Section~\ref{sec:results} reports
the model comparison, the feature-subset ablation, the tuned selector, and the
transfer test. Section~\ref{sec:conclusion} discusses the findings, their
limitations, and directions for future work.

%==============================================================================
\section{Problem Formulation}\label{sec:problem}
%==============================================================================
For a graph $G$ let $\cxp(G)$ and $\cxm(G)$ be the number of CX (two-qubit)
gates required to implement a Trotterization circuit over the Pauli and matching decompositions, respectively. We use CX count as the
cost since two-qubit gates dominate error and runtime on current hardware.
Circuit depth is recorded alongside the gate counts during labeling but is not
modeled here. The selector targets CX count alone, and a multi-objective
formulation is left to future work (Section~\ref{sec:conclusion}).

Define the \emph{signed cost gap}
\begin{equation}
\dcx(G) \;=\; \cxp(G) - \cxm(G),
\end{equation}
so $\dcx(G) > 0$ exactly when Trotterization over the matching decomposition methods requires fewer CX gates than Trotterization over Pauli decomposition. The selection target is the
binary label
\begin{equation}
y(G) \;=\; \mathbf{1}\!\left[\dcx(G) > 0\right],
\end{equation}
with $y=1$ meaning ``use matching'' and $y=0$ meaning ``use Pauli.'' Ties
($\dcx=0$) are excluded from training and evaluation. A model learns the $\dcx=0$
decision surface separating the matching-favorable region of feature space from
the Pauli-favorable region.

%==============================================================================
\section{Dataset}\label{sec:dataset}
%==============================================================================
Our primary testbed is the 11,117 8-vertex connected graph dataset \cite{mckay}, and two datasets containing graphs with between 8 and 256 vertices 
supply a balanced held-out test set. We describe the population first because it
carries the paper's main result.

\subsection{The McKay population (primary)}\label{sec:mckay}
The central dataset is the complete population of all connected graphs on eight
vertices, enumerated by Brendan McKay's \texttt{geng} tool from the \texttt{nauty}
suite~\cite{mckay2014practical}. There are exactly $11{,}117$ such graphs. Because
we take the \emph{entire} population rather than a sample from it, its class
balance and class overlap are properties of the population itself and not
estimated.

Labeling every graph with the code repository of~\cite{atallah2026simulating} gives
$418$ matching-wins ($3.76\%$), $10{,}560$ Pauli-wins, and $139$ ties. Dropping
the ties, as in Section~\ref{sec:problem}, leaves $10{,}978$ modeling rows of
which $418$ are positive, a class imbalance of roughly $25{:}1$ (Pauli to
matching). %This is the setting the rest of the paper
%works in: a complete, exactly-imbalanced population where a model cannot benefit
%from a curated class split.

\subsection{Source families (held-out test)}
Two graph datasets, taken from the companion
work's repository~\cite{atallah2026simulating}, supply the balanced dataset we use
\emph{only} as a held-out test set (Section~\ref{sec:transfer}). No model is
trained on them.

\noindent \textbf{Erd\H{o}s-R\'enyi (ER).} Random $G(n,p)$ graphs, generated by sweeping the
edge probability $p$ from $0.05$ to $0.95$ and drawing graphs at each vertex count
$N\in\{8,16,32,64,128,256\}$. After deduplication the family contains $10{,}981$
graphs, with $1{,}674$, $1{,}800$, $1{,}807$, $1{,}900$, $1{,}900$, and $1{,}900$
graphs at $N=8$, $16$, $32$, $64$, $128$, and $256$, respectively. Their edges
spread across many Hamming distances, which the matching decomposition cannot
exploit, so the Pauli decomposition uses fewer CX gates on all but $0.2\%$ of these
graphs (an imbalance of about $548{:}1$ toward Pauli). We refer to this family as
\emph{Pauli-favorable}.

\noindent \textbf{Structured.} Deterministic counting-path graphs (vertices labeled
$0,1,\dots,N{-}1$ as bitstrings and connected in that order) together with
XOR-permuted, extra-edge, segment-reversed, and perturbed variants. Their edges
cluster at a few Hamming distances, which the matching decomposition compresses
effectively. After deleting duplicate graphs, the family contains $8{,}835$ graphs, with $779$,
$1{,}749$, $1{,}769$, $1{,}746$, $1{,}713$, and $1{,}079$ graphs at $N=8$, $16$,
$32$, $64$, $128$, and $256$, respectively. On this family, Trotterizing over matching
decomposition requires fewer CX gates than Pauli decomposition for the majority of graphs ($75.3\%$), increasingly so at
larger sizes. We refer to it as \emph{matching-favorable}.

\noindent \textbf{Balanced test dataset.} From these two families we form a class-balanced test
set by taking, at each vertex count, an equal number of matching-wins (from the
structured family) and Pauli-wins (from the ER family). The number of each class is
$99$, $329$, $1{,}755$, $1{,}746$, $1{,}713$, and $1{,}079$ at $N=8$, $16$, $32$,
$64$, $128$, and $256$, respectively, giving $198$, $658$, $3{,}510$, $3{,}492$,
$3{,}426$, and $2{,}158$ graphs at those sizes and $13{,}442$ in total. This yields a
$50/50$ class split at every size on which to test the transfer of the model tuned
on the McKay population.

Every dataset is restricted to connected graphs and deduplicated by
Weisfeiler-Leman hash~\cite{hagberg2008exploring} before any labeling or balancing,
so no graph appears twice. %The McKay population is at $N=8$, and the source families
%span the power-of-two vertex counts $N\in\{8,16,32,64,128,256\}$. 
Ground-truth CX
counts are obtained by calling the code repository
of~\cite{atallah2026simulating}, and we do not reimplement the decompositions.

The dataset stops at $N=256$ because, when we fit a power law to the
per-graph labeling times (single core, hardware and software listed in the
\hyperref[sec:repro]{Reproducibility} section), the labeling has complexity approximately $O(N^{3.2})$, which is projected to require about $0.5$ hours per graph at $N=512$ and $4$ hours per
graph at $N=1024$. Labeling even a few hundred graphs at $N=1024$ is therefore a
multi-week computation. Profiling attributes the bulk of that cost to constructing the
Pauli circuit rather than transpiling it, so it is not recoverable by changing
compiler flags. Against this $O(N^{3.2})$ labeling cost, extracting the
decomposition features is far less time intensive (Section~\ref{sec:features}),
however it bounds what we can validate (Section~\ref{sec:conclusion}).

\subsection{The feature set}\label{sec:features}
We describe each graph by twelve features and make no prior assumption about which
of them matter. Ten are \emph{topological}, standard structural properties of the
graph $G=(V,E)$, with vertex set $V$ ($|V|=N$) and edge set $E$, computed with
NetworkX~\cite{hagberg2008exploring}. We use the notation 
$d_1,\dots,d_N$ for the vertex degrees, $\bar d$ for their mean, $A$ for the
adjacency matrix, $D=\mathrm{diag}(d_1,\dots,d_N)$ for the diagonal degree matrix,
and $L=D-A$ for the graph Laplacian.
Writing $m=|E|$, the ten features are the following.
\begin{itemize}
\item \emph{Edge density:} $d = \dfrac{2m}{N(N-1)}$, the ratio of the number of edges
to the maximum possible for a simple graph on $N$ vertices.
\item \emph{Average}, \emph{maximum}, and \emph{minimum degree:} $\bar d$,
$\max_i d_i$, and $\min_i d_i$.
\item \emph{Degree variance:} $\dfrac{1}{N}\sum_i (d_i-\bar d)^2$, which measures how
uneven the degrees are.
\item \emph{Maximum matching size:}, the number of edges in a maximum-cardinality
\emph{matching}, that is, a largest subset of $E$ in which no vertex is incident to
more than one edge.
\item \emph{Spectral gap:} $\lambda_2(L)$, the second-smallest eigenvalue of the
Laplacian, also known as the algebraic connectivity. $L = D-A$ where $A$ is the adjacency matrix of the graph and $D$ is the diagonal matrix that encodes the degree of each vertex.
\item \emph{Average clustering coefficient:} $C = \dfrac{1}{N}\sum_i c_i$, where the
local coefficient $c_i = \dfrac{2\,T_i}{d_i(d_i-1)}$ is the fraction of possible
triangles through vertex $i$ that are present and $T_i$ is the number of triangles
containing $i$.
\item \emph{Triangle count:} $\tfrac{1}{3}\sum_i T_i$, the number of triangles in
$G$.
\item \emph{Diameter:} $\max_i \varepsilon(i)$, the maximum eccentricity, where the
eccentricity $\varepsilon(i)=\max_j \mathrm{dist}(i,j)$ is the greatest shortest-path
distance from vertex $i$ to any other.
\end{itemize}
Fitting a power law to the per-graph time for the full set of ten features, as we
do for the labeling cost, gives about $O(N^{1.3})$ over
$N\in\{8,16,32,64,128,256\}$, orders of
magnitude below the $O(N^{3.2})$ cost of labeling. Under the same fit, the
fastest-growing individual features are the diameter (an all-pairs shortest-path
computation) at about $O(N^{1.8})$ and the maximum matching at about $O(N^{1.6})$,
which set the scaling at the larger sizes, while the remaining features.

The remaining two features are \emph{decomposition-derived}: the number of Pauli
terms $n_{\mathrm{Pauli}}$ in the Pauli decomposition, and the number of matchings
$n_{\mathrm{match}}$ in the matching decomposition. Both are obtained by
constructing the decomposition and counting its terms, without building or
transpiling the Trotter circuit, which is the expensive part of labeling. We compute
$n_{\mathrm{Pauli}}$ with the tensorized Pauli decomposition
of~\cite{hantzko2023tensorized} as implemented in Qiskit's
\lstinline[style=pystyle]|SparsePauliOp.from_operator|, which returns the full set of Pauli coefficients without enumerating the terms one at a time. Although the adjacency
matrix defines the Hamiltonian, its expansion in the Pauli basis, and hence
$n_{\mathrm{Pauli}}$, depends on the binary labeling that assigns vertices to qubit
indices, so $n_{\mathrm{Pauli}}$ is a property of the labeled Hamiltonian rather than
an invariant of the unlabeled graph. We use a single fixed vertex labeling for every
graph, so $n_{\mathrm{Pauli}}$ is well defined throughout, with the understanding that
a different labeling could give a different count.

%We then count the
% terms whose coefficient is non-negligible, as in Listing~\ref{lst:pauli}, at a cost
% near $O(N^{1.2})$ per graph.
% \begin{lstlisting}[style=pystyle,caption={Computing $n_{\mathrm{Pauli}}$ from the
% tensorized Pauli decomposition.},label={lst:pauli},captionpos=b]
% spo = SparsePauliOp.from_operator(Operator(H))
% n_pauli = int((abs(spo.coeffs) > 1e-12).sum())
% \end{lstlisting}
We obtain $n_{\mathrm{match}}$ from the reference matching-decomposition
implementation~\cite{atallah2026simulating}, which partitions the graph's edges into
matchings of equal Hamming distance using its compression-aware heuristic, retaining
the best of $30$ randomized trials, and we take $n_{\mathrm{match}}$ to be the number
of matchings in that partition, computed at a cost near $O(N^{1.5})$ per graph.

These two counts are the features we expect to matter, since they are
quantities derived from the decompositions, but they are surrogates
for the true CX counts. Computing all twelve
features costs about $O(N^{1.5})$ per graph, dominated by the matching count,
whereas building and transpiling both circuits to obtain the true CX counts costs
$O(N^{3.2})$, so the features become far cheaper than the label as the graphs grow
(Table~\ref{tab:cost}). The gap is already large at the sizes we test. At $N=256$,
$n_{\mathrm{Pauli}}$, which is the single most predictive feature (Section~\ref{sec:eda}),
takes $0.003$\,s per graph, against $157$\,s to build and transpile the circuits for the
true label, a factor of about $50{,}000$, and the full twelve-feature set still costs
only $0.55$\,s. Predicting the label from
these features therefore means predicting an expensive quantity from relatively cheap features,
rather than avoiding the decompositions altogether. Whether these counts actually
suffice is settled empirically in Section~\ref{sec:eda}.

\begin{table}[t]
\caption{Per-graph cost of each feature block against labeling, with empirical
scaling fit over $N\in\{8,16,32,64,128,256\}$ (single core, hardware in the
\hyperref[sec:repro]{Reproducibility} section).}
\label{tab:cost}
\centering
\setlength{\tabcolsep}{5pt}
\begin{tabular}{lcc}
\toprule
operation & scaling in $N$ & per graph at $N{=}256$ \\
\midrule
$n_{\mathrm{Pauli}}$ (tensorized Pauli count) & $O(N^{1.2})$ & $0.003$\,s \\
ten topological features & $O(N^{1.3})$ & $0.05$\,s \\
$n_{\mathrm{match}}$ (matching count) & $O(N^{1.5})$ & $0.50$\,s \\
\midrule
all $12$ features & $O(N^{1.5})$ & $\mathbf{0.55}$\,s \\
build $+$ transpile (labeling) & $O(N^{3.2})$ & $\mathbf{157}$\,s \\
\bottomrule
\end{tabular}
\end{table}

\subsection{Split protocol}
Using the complete population means that we label and analyze every eight-vertex
graph, but the models themselves are still fit and evaluated on disjoint subsets of
it, so that reported performance is always measured on graphs the model did not see
during training. We split the McKay population into a $64/16/20$ train, validation,
and test partition, stratified by class, giving $7{,}025$ training, $1{,}757$
validation, and $2{,}196$ test graphs, and each result on the McKay population is
averaged over five such random splits. The balanced dataset is never used for
training. It is drawn from the ER and structured families, balanced within each
vertex count, and used in full as a held-out test set. Table~\ref{tab:data} lists
the datasets and this split.

\begin{table*}[t]
\caption{Datasets. ``win \%'' is the percentage of graphs where matching is cheaper,
with ties ($\dcx=0$) excluded from all counts. The train/val/test columns give the
$64/16/20$ stratified split applied to the McKay population. The balanced dataset
is used only as a held-out test set and the ER and structured families only as
its two sources, so these three receive no such split (marked n/a).}
\label{tab:data}
\centering
\setlength{\tabcolsep}{6pt}
\begin{tabular}{lrrrrrrr}
\toprule
dataset & graphs & match-wins & Pauli-wins & win \% & train & val & test \\
\midrule
McKay $N{=}8$   & 10{,}978 & 418     & 10{,}560 & 3.8\%  & 7{,}025 & 1{,}757 & 2{,}196 \\
Erd\H{o}s-R\'enyi (ER) & 10{,}981 & 20 & 10{,}961 & 0.2\%  & n/a & n/a & n/a \\
structured      &  9{,}026 & 6{,}879 & 2{,}147 & 76.2\% & n/a & n/a & n/a \\
balanced        & 13{,}442 & 6{,}721 & 6{,}721 & 50\%   & n/a & n/a & 13{,}442 \\
\bottomrule
\end{tabular}
\end{table*}

%==============================================================================
\section{Data Analysis}\label{sec:eda}
%==============================================================================
Before training any classifier, we examine the population to establish which of the
twelve features carry information about the label and whether the two classes are
separable in feature space. Throughout this section we call a graph a
\emph{matching-win} if the matching decomposition compiles it to fewer CX gates than
the Pauli decomposition, and a \emph{Pauli-win} otherwise. As Table~\ref{tab:data}
records, the population contains only $418$ matching-wins against $10{,}560$
Pauli-wins, an imbalance of roughly $25{:}1$.

\begin{figure}[t]
\centering
\includegraphics[width=\columnwidth]{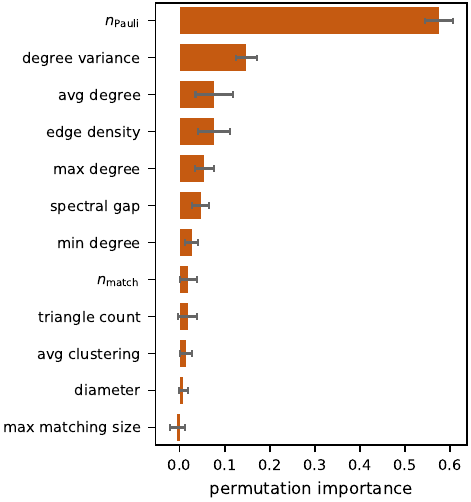}
\caption{Permutation feature importance on the training population, measured as the
mean drop in predictive performance when each feature is shuffled (five seeds).}
\label{fig:importance}
\end{figure}

We first try to determine how much each feature matters using \emph{permutation importance}: for
a model fitted on all twelve features, each feature is shuffled in turn and the
resulting drop in predictive performance measures how much the model relies on it.
Fig.~\ref{fig:importance} reports this drop for each feature, averaged over twenty
permutations and five train/test splits.

As Fig.~\ref{fig:importance} shows, the Pauli term count $n_{\mathrm{Pauli}}$ is by
far the most important feature, at $0.575$, nearly four times the next feature. The
strongest topological feature is the degree variance, at $0.148$.
The remaining features are all much smaller: average degree $0.077$, edge density
$0.075$, maximum degree $0.055$, spectral gap $0.047$, minimum degree $0.027$, the
matching count $n_{\mathrm{match}}$ $0.019$, triangle count $0.018$, average
clustering $0.014$, diameter $0.008$, and maximum matching size $-0.005$. The value
plotted for each feature is its permutation importance, the drop in test MCC when that
feature is shuffled, not the feature itself, so the small negative entry for the
maximum matching size means only that shuffling it left performance essentially
unchanged, with the tiny negative sign a sampling artifact of a feature the model does
not use, rather than any negative quantity of the feature.

\begin{figure}[t]
\centering
\includegraphics[width=0.85\columnwidth]{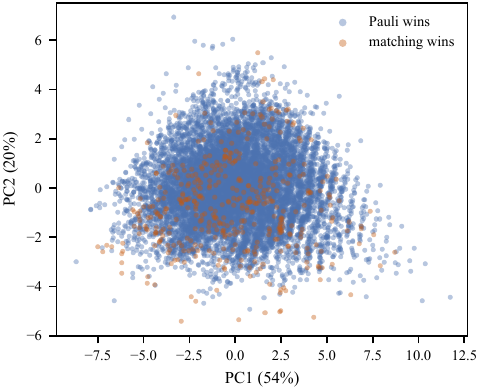}
\caption{First two principal components of the twelve standardized features on the
McKay population, with each graph colored by class. The matching-wins overlap the
Pauli-wins rather than separating from them, so the two classes are not separated in
these two principal directions.}
\label{fig:pca}
\end{figure}

We next examine how the two classes sit in feature space. Fig.~\ref{fig:pca} projects
the standardized feature vectors onto their first two principal components, the two
directions of greatest variance. The matching-wins are dispersed among the Pauli-wins
rather than occupying a distinct region, so the classes are not separated along these
two directions. This does not by itself rule out a linear boundary in the full
twelve-dimensional space, but the linear models trained on all twelve features in
Section~\ref{sec:results} do not find one, consistent with the overlap seen here.

\begin{figure}[t]
\centering
\includegraphics[width=\columnwidth]{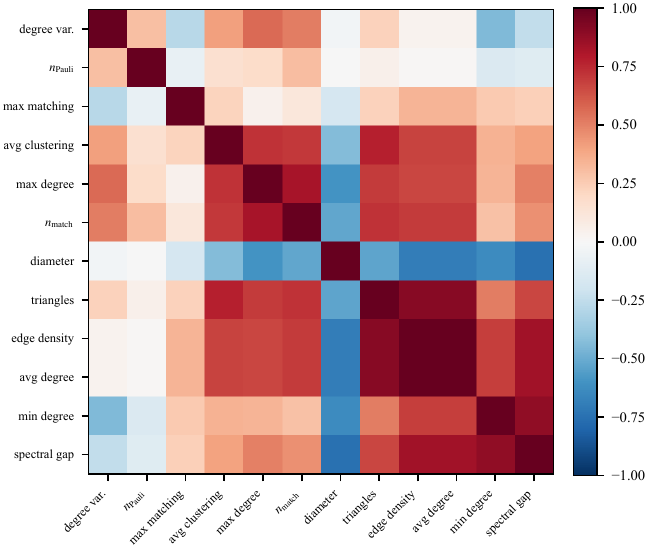}
\caption{Pairwise Pearson correlation between the twelve features on the McKay
population, with features ordered by hierarchical clustering so that correlated
groups appear adjacent.}
\label{fig:corr}
\end{figure}

Finally, Fig.~\ref{fig:corr} shows the pairwise Pearson correlation between the
features, where for two features $x$ and $y$ the coefficient
\begin{equation}
\rho_{xy} = \frac{\sum_i (x_i-\bar x)(y_i-\bar y)}
{\sqrt{\sum_i (x_i-\bar x)^2}\,\sqrt{\sum_i (y_i-\bar y)^2}}
\end{equation}
measures their linear dependence, ranging from $-1$ to $1$, where $x_i$ and $y_i$ are
the values of the two features for graph $i$ and $\bar x$ and $\bar y$ are the means of the features. The topological features 
are strongly correlated with one another (mean pairwise $|\rho|\approx 0.51$), and
$n_{\mathrm{match}}$ lies within this correlated group (mean $|\rho|\approx 0.55$,
strongest with maximum degree at $0.83$), behaving as one more structural feature.
Only $n_{\mathrm{Pauli}}$ stands apart, nearly uncorrelated with the rest (mean
$|\rho|\approx 0.12$), and supplies information the other features do not.

%==============================================================================
\section{Model and Evaluation Protocol}\label{sec:model}
%==============================================================================
In this section, we discuss the evaluation metrics for each model, as well as the models we used.
\subsection{Metrics}
Taking matching as the positive class, every prediction is one of four outcomes,
depending on whether the predicted class matches the true one. The confusion matrix
collects the counts of these outcomes over the test set, with true classes on the
rows and predicted classes on the columns,
\begin{equation*}
\begin{array}{r|cc}
 & \text{pred. matching} & \text{pred. Pauli} \\
\hline
\text{true matching} & T_p & F_n \\
\text{true Pauli}    & F_p & T_n \\
\end{array}
\end{equation*}
so a true positive ($T_p$) or false negative ($F_n$) is a matching-wins graph
predicted as matching or as Pauli respectively, and a true negative ($T_n$) or false
positive ($F_p$) is a Pauli-wins graph predicted as Pauli or as matching. Our two threshold-fixed metrics are
\begin{equation*}
\text{accuracy} = \frac{T_p+T_n}{T_p+T_n+F_p+F_n}, \qquad
\text{F1} = \frac{2 T_p}{2 T_p+F_p+F_n},
\end{equation*}
together with the Matthews correlation coefficient (MCC)~\cite{matthews1975comparison}
\begin{equation*}
\mathrm{MCC} = \frac{T_p T_n - F_p F_n}
{\sqrt{(T_p{+}F_p)(T_p{+}F_n)(T_n{+}F_p)(T_n{+}F_n)}},
\end{equation*}
which ranges from $-1$ to $1$ and equals $0$ for a model no better than random. We
also report two threshold-independent measures obtained by varying the decision
threshold, the areas under the receiver-operating-characteristic and
precision-recall curves (ROC-AUC and PR-AUC). The four class-specific rates
(precision, recall, specificity, NPV) are defined in Appendix~\ref{sec:appendix-metrics}.

MCC and PR-AUC are our primary metrics because they remain informative under the
strong class imbalance of the McKay population, whereas accuracy does
not~\cite{chicco2020advantages,saito2015precision}. With matching-wins outnumbered
roughly $25$ to $1$, a trivial model that always predicts Pauli already attains
about $96\%$ accuracy, so a high accuracy does not indicate that the rare
matching-wins are being found, whereas MCC and PR-AUC only reward a model that
correctly identifies the minority class. We also examine precision and recall
together, because a model can raise its recall, the fraction of matching-wins it
detects, simply by predicting matching more often, at the cost of a lower
precision, the fraction of its matching predictions that are correct. For this
reason the result Tables~\ref{tab:ladder}--\ref{tab:tuned} report accuracy, F1, MCC,
and the two areas ROC-AUC and PR-AUC, with the remaining class-specific rates
given in the appendix.

\subsection{Models}\label{sec:baselines}
Rather than commit to a single model in advance, we train twelve classifiers on the
full twelve-feature set and compare them under a common protocol. The twelve models were
chosen to span a range of decision-boundary shapes, so that if the signal is best
captured by a particular kind of model, one of them will find it. They are logistic
regression, Gaussian naive Bayes (linear boundaries), decision trees of depth
$1$, $2$, and $3$ (axis-aligned splits), a $k$-nearest-neighbor classifier with
$k=5$, a support vector machine with a radial basis function kernel (RBF-SVM)
(local and kernel boundaries), a multilayer perceptron (a feedforward neural network
whose activation function is linear), and four tree
ensembles: a random forest~\cite{breiman2001random}, gradient boosting,
XGBoost~\cite{chen2016xgboost}, and LightGBM~\cite{ke2017lightgbm}. All use the scikit-learn
implementations~\cite{pedregosa2011scikit}. The models whose predictions depend on
feature scale (logistic regression, naive Bayes, $k$-NN, the RBF-SVM, and the ANN)
are given standardized features, with the standardizer fit on the training rows
only, while the scale-invariant tree-based models receive the raw features, so that
every model sees the same information in the form its algorithm requires.

We first compare the twelve models using their default settings. We then tune each one
by a per-model grid search on the McKay population over the same feature set and
report its best configuration. Every reported number is the mean over five random
train/test splits.

%==============================================================================
\section{Results}\label{sec:results}
%==============================================================================
This section reports three experiments. We first evaluate the untuned models
on the full feature set. We then tune each model and identify the best
configuration, which is a small artificial neural network. Finally, we take that tuned
model and, without retraining it, measure how well it predicts the cheaper
decomposition on the balanced held-out set of larger graphs.

\subsection{Comparison of models}\label{sec:ladder}
Table~\ref{tab:ladder} reports the models on the full twelve-feature set,
sorted by MCC. Gradient boosting is best, at MCC $0.569$ (precision $0.766$, recall
$0.445$, PR-AUC $0.636$), followed by logistic regression at MCC $0.554$ (precision
$0.742$, recall $0.436$, PR-AUC $0.653$). The neural and tree-ensemble models follow
closely, between MCC $0.506$ and $0.515$.

The results also illustrate why we use MCC and PR-AUC as success metrics rather than accuracy on
this imbalanced data. Two models achieve a nonzero MCC by very different means. The
RBF-SVM predicts matching for almost no graphs, so its few matching predictions are
nearly all correct (precision $0.96$) but it detects only $18\%$ of the true
matching-wins (recall $0.183$). Gaussian naive Bayes does the opposite, predicting
matching often enough to detect $33\%$ of the matching-wins (recall $0.329$), but
only $27\%$ of its matching predictions are correct (precision $0.27$), which brings its MCC down
to $0.266$. The lowest-ranked model is the depth-one decision tree, which splits on
a single feature once and performs barely above random (MCC $0.082$). Because MCC
and PR-AUC penalize both of these failure modes, they order the models in a way that
accuracy, which stays near $0.96$ throughout, cannot.

\begin{table*}[t]
\caption{Model comparison on the McKay population, full feature set, sorted by MCC
(mean $\pm$ std over five seeds). Remaining metrics in
Table~\ref{tab:ladder-appendix}.}
\label{tab:ladder}
\centering
\setlength{\tabcolsep}{6pt}
\begin{tabular}{lccccc}
\toprule
model & acc. & F1 & MCC & ROC-AUC & PR-AUC \\
\midrule
\textbf{gradient boosting} & \pms{0.973}{0.001} & \pms{0.558}{0.034} & \pms{\mathbf{0.569}}{0.023} & \pms{0.957}{0.01} & \pms{0.636}{0.015} \\
logistic & \pms{0.972}{0.001} & \pms{0.546}{0.04} & \pms{0.554}{0.03} & \pms{0.961}{0.01} & \pms{0.653}{0.032} \\
LightGBM & \pms{0.969}{0.001} & \pms{0.519}{0.036} & \pms{0.515}{0.031} & \pms{0.952}{0.009} & \pms{0.574}{0.021} \\
random forest & \pms{0.97}{0.001} & \pms{0.5}{0.054} & \pms{0.513}{0.041} & \pms{0.939}{0.015} & \pms{0.599}{0.026} \\
ANN & \pms{0.969}{0.003} & \pms{0.515}{0.043} & \pms{0.512}{0.045} & \pms{0.949}{0.012} & \pms{0.563}{0.055} \\
XGBoost & \pms{0.968}{0.001} & \pms{0.516}{0.025} & \pms{0.506}{0.023} & \pms{0.947}{0.01} & \pms{0.556}{0.02} \\
decision tree ($d{=}3$) & \pms{0.968}{0.001} & \pms{0.446}{0.101} & \pms{0.471}{0.067} & \pms{0.893}{0.044} & \pms{0.471}{0.041} \\
SVM (RBF) & \pms{0.968}{0.001} & \pms{0.307}{0.04} & \pms{0.411}{0.037} & \pms{0.931}{0.023} & \pms{0.61}{0.034} \\
$k$-NN & \pms{0.967}{0.003} & \pms{0.327}{0.069} & \pms{0.38}{0.075} & \pms{0.827}{0.034} & \pms{0.362}{0.042} \\
decision tree ($d{=}2$) & \pms{0.966}{0.001} & \pms{0.289}{0.092} & \pms{0.367}{0.049} & \pms{0.882}{0.044} & \pms{0.368}{0.023} \\
naive Bayes & \pms{0.94}{0.006} & \pms{0.296}{0.039} & \pms{0.266}{0.041} & \pms{0.901}{0.02} & \pms{0.239}{0.03} \\
decision tree ($d{=}1$) & \pms{0.961}{0.001} & \pms{0.086}{0.172} & \pms{0.082}{0.164} & \pms{0.744}{0.038} & \pms{0.257}{0.028} \\
\bottomrule
\end{tabular}
\end{table*}

\subsection{Hyperparameter tuning}\label{sec:tuning}
The comparison of untuned models establishes that the full feature set is
learnable. We now tune each model by a per-model grid search over the McKay
population and report the best configuration of each in Table~\ref{tab:tuned},
sorted by MCC, with the selected hyperparameters listed in Table~\ref{tab:config}.
The top models are closely bunched. An artificial neural network (ANN) with a single
hidden layer of $32$ nodes (tanh activation, $\ell_2$ penalty $\alpha=10^{-4}$) has the
highest mean MCC at $0.593$, but gradient boosting follows at $0.583$ and the next
several models trail by comparable margins, all within the seed-to-seed variation of a
few hundredths. We do not read this ordering as evidence that one architecture is
inherently better. We simply adopt the ANN as the selected model for the remaining
experiments, reaching MCC $0.593$ at recall $0.476$ and precision $0.774$.

\begin{figure*}[t]
\centering
\includegraphics[width=0.99\textwidth]{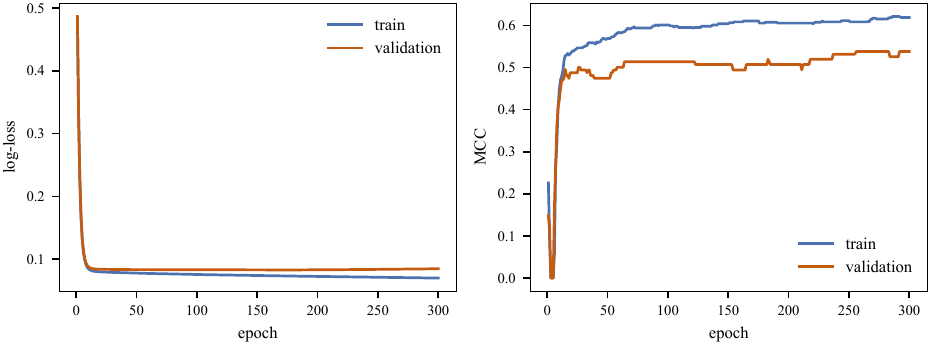}
\caption{Training and validation convergence of the tuned ANN on the McKay split
(seed~$0$): log-loss (left) and MCC (right) versus epoch. Train and validation
track together, indicating the single $32$-unit layer generalizes rather than
memorizes.}
\label{fig:conv}
\end{figure*}

Within the ANN family, the lowest-capacity network is preferred. Every deeper or wider
ANN in the grid scored below the single hidden layer of $32$ nodes. With only $418$
matching-wins in the whole population, the additional parameters of a larger network
have too little signal to fit and instead memorize the training data. Fig.~\ref{fig:conv} confirms that the chosen model does not
suffer from this problem. Over the course of training, its training and validation
log-loss decrease together and its training and validation MCC increase together to
a small and stable gap, whereas an over-parameterized model would show the
validation curves diverging from the training curves.

The precision and recall values in Table~\ref{tab:tuned} again distinguish the
models. The random forest reaches MCC $0.558$, but only by predicting matching often
enough to detect $70\%$ of the matching-wins (recall $0.698$) at a precision of
$0.48$, and the SVM and logistic regression push recall above $0.9$ with precision
near $0.22$. The ANN and the tuned tree ensembles are the models that combine a high MCC with
balanced precision and recall, which is what motivates adopting one of them.

\begin{table*}[t]
\caption{Per-model tuned sweep on the McKay population, best configuration each,
sorted by MCC. Remaining metrics in Table~\ref{tab:tuned-appendix}.}
\label{tab:tuned}
\centering
\setlength{\tabcolsep}{6pt}
\begin{tabular}{lccccc}
\toprule
model & acc. & F1 & MCC & ROC-AUC & PR-AUC \\
\midrule
\textbf{ANN} & \pms{0.974}{0.001} & \pms{0.586}{0.045} & \pms{\mathbf{0.593}}{0.036} & \pms{0.961}{0.009} & \pms{0.659}{0.026} \\
gradient boosting & \pms{0.974}{0.001} & \pms{0.574}{0.017} & \pms{0.583}{0.013} & \pms{0.955}{0.011} & \pms{0.633}{0.026} \\
random forest & \pms{0.959}{0.003} & \pms{0.568}{0.029} & \pms{0.558}{0.032} & \pms{0.95}{0.01} & \pms{0.628}{0.029} \\
XGBoost & \pms{0.971}{0.002} & \pms{0.544}{0.026} & \pms{0.543}{0.025} & \pms{0.956}{0.007} & \pms{0.596}{0.016} \\
LightGBM & \pms{0.971}{0.001} & \pms{0.53}{0.033} & \pms{0.536}{0.026} & \pms{0.957}{0.009} & \pms{0.603}{0.026} \\
decision tree & \pms{0.937}{0.005} & \pms{0.436}{0.021} & \pms{0.431}{0.02} & \pms{0.802}{0.008} & \pms{0.421}{0.026} \\
SVM (RBF) & \pms{0.878}{0.005} & \pms{0.364}{0.014} & \pms{0.419}{0.018} & \pms{0.958}{0.008} & \pms{0.623}{0.042} \\
logistic & \pms{0.877}{0.005} & \pms{0.361}{0.018} & \pms{0.416}{0.025} & \pms{0.959}{0.01} & \pms{0.627}{0.035} \\
$k$-NN & \pms{0.966}{0.001} & \pms{0.385}{0.029} & \pms{0.404}{0.024} & \pms{0.783}{0.033} & \pms{0.316}{0.038} \\
naive Bayes & \pms{0.94}{0.006} & \pms{0.296}{0.039} & \pms{0.267}{0.041} & \pms{0.901}{0.02} & \pms{0.239}{0.03} \\
\bottomrule
\end{tabular}
\end{table*}

\begin{table*}[t]
\caption{Best configuration selected for each model by the per-model grid search
on the McKay population (maximizing MCC), sorted by MCC.}
\label{tab:config}
\centering
\setlength{\tabcolsep}{6pt}
\begin{tabular}{@{}l p{6.2cm} @{\hspace{2em}} l p{6.2cm}@{}}
\toprule
model & best hyperparameters & model & best hyperparameters \\
\midrule
\textbf{ANN} & one $32$-unit hidden layer, tanh, $\alpha=10^{-4}$, lr $10^{-3}$ &
decision tree & entropy split, unbounded depth, min.\ leaf $5$ \\
gradient boosting & $400$ trees, depth $2$, learning rate $0.1$, subsample $0.8$ &
SVM (RBF) & $C=10$, $\gamma=0.01$ \\
random forest & $400$ trees, unbounded depth, min.\ leaf $5$ &
logistic & $\ell_2$ penalty, $C\approx3.16$, lbfgs solver \\
XGBoost & $400$ trees, depth $3$, learning rate $0.1$, subsample $1$ &
$k$-NN & $3$ neighbors, uniform weights, Euclidean \\
LightGBM & $400$ trees, $15$ leaves, learning rate $0.02$, min.\ child $20$ &
naive Bayes & Gaussian, variance smoothing $10^{-3}$ \\
\bottomrule
\end{tabular}
\end{table*}

\subsection{Model behavior}\label{sec:behavior}
\begin{figure*}[t]
\centering
\begin{minipage}{\columnwidth}
\centering
\includegraphics[width=\textwidth]{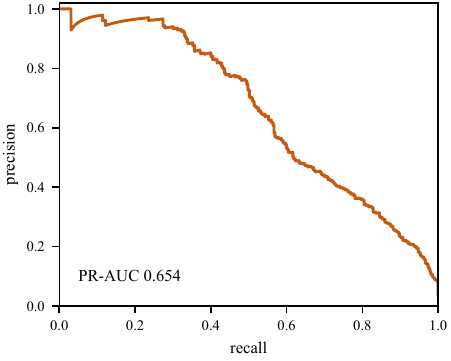}
\subcaption{Precision-Recall}
\label{fig:pr}
\end{minipage}
\hfill
\begin{minipage}{\columnwidth}
\centering
\includegraphics[width=\textwidth]{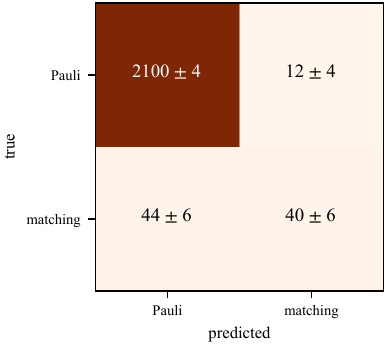}
\subcaption{Confusion matrix (per-fold mean $\pm$ std)}
\label{fig:cm}
\end{minipage}
\caption{Tuned ANN evaluated on the held-out McKay test graphs, averaged over five
random splits. In (a), PR-AUC is the area under the precision-recall curve. A value close to zero indicates a model does no better than random guessing. In (b),
each cell counts the test graphs of a given true class (rows) assigned to a given
predicted class (columns), with the diagonal holding the correct predictions.}
\label{fig:eval}
\end{figure*}

Fig.~\ref{fig:pr} shows the precision-recall curve of the tuned ANN on the held-out
test graphs, which is the preferred way to summarize a classifier under strong class
imbalance~\cite{saito2015precision}. The model keeps precision above $0.8$, 
recall of approximately $0.4$, and attains an average precision of $0.654$, far above the
$0.038$ that a random guess would achieve. Fig.~\ref{fig:cm} gives the confusion
matrix of Section~\ref{sec:model} for the model's predictions. Each graph is scored only
when it is in the held-out test set, never during training, and each cell reports the mean and standard
deviation over the five random splits (each test set is about $20\%$ of the
population, so roughly $84$ matching-wins and $2{,}112$ Pauli-wins). On average the
model correctly identifies $40\pm6$ of the matching-wins while making only $12\pm4$
incorrect matching predictions against $2{,}100\pm4$ correctly identified Pauli-wins,
a high-precision, moderate-recall operating point.

\subsection{Feature ablation}\label{sec:ablation}
The importance ranking of Fig.~\ref{fig:importance} identifies which features a
single model relies on, but not how the features combine, in particular whether two
important features supply the same information or different information. To test
this, we retrain the tuned ANN on a small set of targeted feature subsets, keeping the training and evaluation protocol fixed so
that only the feature set changes. The subsets are the ten topological features
alone, the Pauli term count alone, the two decomposition counts together, the Pauli
term count paired with the degree variance, and the full twelve-feature set.
Table~\ref{tab:ablation} reports the result, from which three conclusions follow.

First, the topological features alone cannot predict which decomposition requires fewer CX gates.
Restricted to the ten topological features, the ANN reaches only MCC $0.062$ at
recall $0.01$, meaning it predicts matching for almost no graphs and so fails to
identify the graphs on which matching would use fewer CX gates. This is the outcome
anticipated by the class overlap of Section~\ref{sec:eda}. Second, the Pauli
term count is the single most useful feature but is not sufficient on its own. Using
$n_{\mathrm{Pauli}}$ alone reaches MCC $0.423$, and adding $n_{\mathrm{match}}$
raises it only to $0.433$. Third, the largest improvement comes from combining the Pauli term count
with the degree variance results in a MCC $0.574$, which recovers
most of the gap to the full twelve-feature set at $0.593$ and raises recall from
$0.293$ to $0.448$ and PR-AUC from $0.452$ to $0.637$. These two features are
complementary: $n_{\mathrm{Pauli}}$ measures the size of the Pauli expansion, while
the degree variance reflects how the matching decomposition can group the edges. As
expected, accuracy and specificity stay between $0.96$ and $0.98$ for every subset,
including the topology-only model that predicts almost nothing correctly, confirming
that on this imbalanced data these two metrics cannot separate a useful model from a
trivial one.

The final row of Table~\ref{tab:ablation} tests the simplest possible alternative to
a learned model: a rule that thresholds the Pauli term count and predicts one
decomposition below the threshold and the other above~\cite{holte1993very}. The
natural expectation is that a large Pauli expansion should require more CX gates and thus should trigger a
switch to matching, so that the rule predicts matching when $n_{\mathrm{Pauli}}>t$. We
therefore let the training data choose both the threshold and the direction. The data
overturns this expectation and selects the opposite rule, predicting matching when
$n_{\mathrm{Pauli}}\le t$. The reason is that $n_{\mathrm{Pauli}}$ tracks graph
complexity rather than the cost of the matching. Matching-wins have a mean
$n_{\mathrm{Pauli}}$ of $12.1$ against $19.1$ for Pauli-wins, because a small Pauli
count marks a sparse graph on which the matching decomposition uses even fewer CX
gates, whereas a large Pauli count marks a dense graph on which the number of
matchings grows faster and the Pauli route stays cheaper despite its larger term
count. We choose $t$ on the training split so as to maximize MCC, then evaluate the
resulting rule on the test split. At $N=8$, where $n_{\mathrm{Pauli}}$ ranges from $5$
to $28$, the selected threshold is $t=11$ on three of the five splits and $t=12$ on
the other two. This
rule reaches MCC $0.466$, better than the tuned ANN given only
$n_{\mathrm{Pauli}}$ but well below the $0.593$ of the same model on the full feature set. Unlike a trained
classifier, it outputs only a hard matching-or-Pauli decision rather than a
probability, so ROC-AUC and PR-AUC, which require ranking graphs by predicted
probability, are undefined for it.

\begin{table*}[t]
\caption{Feature-subset ablation for the tuned ANN on the McKay population
(mean $\pm$ std over five seeds). The last row is a one-rule threshold on
$n_{\mathrm{Pauli}}$, which emits no score ($\cdot$). Remaining metrics in
Table~\ref{tab:ablation-appendix}.}
\label{tab:ablation}
\centering
\setlength{\tabcolsep}{6pt}
\begin{tabular}{lccccc}
\toprule
feature subset & acc. & F1 & MCC & ROC-AUC & PR-AUC \\
\midrule
topological only ($10$) & \pms{0.962}{0} & \pms{0.019}{0.009} & \pms{0.062}{0.035} & \pms{0.726}{0.017} & \pms{0.166}{0.025} \\
$n_{\mathrm{Pauli}}$ only & \pms{0.967}{0.001} & \pms{0.402}{0.051} & \pms{0.423}{0.038} & \pms{0.938}{0.01} & \pms{0.452}{0.037} \\
$n_{\mathrm{Pauli}}+n_{\mathrm{match}}$ & \pms{0.967}{0.002} & \pms{0.415}{0.075} & \pms{0.433}{0.057} & \pms{0.946}{0.011} & \pms{0.5}{0.042} \\
$n_{\mathrm{Pauli}}+$ degree var. & \pms{0.974}{0.001} & \pms{0.564}{0.031} & \pms{0.574}{0.022} & \pms{0.961}{0.01} & \pms{0.637}{0.033} \\
\textbf{full set ($12$)} & \pms{0.974}{0.001} & \pms{0.586}{0.045} & \pms{\mathbf{0.593}}{0.036} & \pms{0.961}{0.009} & \pms{0.659}{0.026} \\
\midrule
threshold on $n_{\mathrm{Pauli}}$ & \pms{0.958}{0.005} & \pms{0.484}{0.03} & \pms{0.466}{0.032} & $\cdot$ & $\cdot$ \\
\bottomrule
\end{tabular}
\end{table*}

\subsection{Transfer to larger graphs}\label{sec:transfer}
The tuned ANN is fit once on the McKay population and never sees the source families
during training. To test how far its decision transfers, we apply the frozen model
to the balanced held-out dataset (Section~\ref{sec:dataset}) at each vertex size.
Table~\ref{tab:persize} reports the result, which improves with graph size. At
$N=8$, the training size but on graphs from a different distribution, the model
reaches MCC $0.785$, accuracy $0.89$, and PR-AUC $0.961$, and at $N=16$ it reaches
MCC $0.711$ at accuracy $0.852$. From $N=32$ onward the two classes are almost
perfectly separable and the model is nearly exact, with MCC $0.993$ at $N=32$ and
$1.0$ at $N\ge64$. A model tuned only on eight-vertex graphs therefore predicts the
cheaper decomposition on graphs an order of magnitude larger that it was never
trained on.

This transfer result should be read with one caveat. The balanced test set draws its
two classes from two different graph families, since the ER family is almost entirely
Pauli-wins ($0.2\%$ matching) and the structured family is mostly matching-wins
($76.2\%$), so on this test set predicting matching is strongly correlated with a
graph coming from the structured family. Part of the near-perfect separation at large
$N$ may therefore reflect a distinction between these two families rather than a
decomposition boundary that transfers to arbitrary graphs. Testing on further graph
families, ideally ones that mix both classes within a single family, is needed to
separate the two.

\begin{table*}[t]
\caption{Tuned ANN tested per vertex size on the balanced held-out dataset (frozen
model, mean $\pm$ std over five seeds). Remaining metrics in
Table~\ref{tab:persize-appendix}.}
\label{tab:persize}
\centering
\setlength{\tabcolsep}{6pt}
\begin{tabular}{rrccccc}
\toprule
$n$ & graphs & acc. & F1 & MCC & ROC-AUC & PR-AUC \\
\midrule
8 & 198 & \pms{0.89}{0.041} & \pms{0.887}{0.044} & \pms{0.785}{0.08} & \pms{0.962}{0.035} & \pms{0.961}{0.037} \\
16 & 658 & \pms{0.852}{0.023} & \pms{0.862}{0.022} & \pms{0.711}{0.046} & \pms{0.933}{0.01} & \pms{0.924}{0.011} \\
32 & 3{,}510 & \pms{0.996}{0.002} & \pms{0.996}{0.002} & \pms{0.993}{0.005} & \pms{0.996}{0.003} & \pms{0.994}{0.005} \\
64 & 3{,}492 & \pms{1}{0} & \pms{1}{0} & \pms{1}{0} & \pms{1}{0} & \pms{1}{0} \\
128 & 3{,}426 & \pms{1}{0} & \pms{1}{0} & \pms{1}{0} & \pms{1}{0} & \pms{1}{0} \\
256 & 2{,}158 & \pms{1}{0} & \pms{1}{0} & \pms{1}{0} & \pms{1}{0} & \pms{1}{0} \\
\bottomrule
\end{tabular}
\end{table*}

%==============================================================================
\section{Discussion and Conclusion}\label{sec:conclusion}
%==============================================================================
In this work, we train a range of machine learning models on the complete population of
all $11{,}117$ connected eight-vertex graphs to determine whether one can learn when
the Pauli decomposition or the matching decomposition requires fewer CX gates for
approximating CTQW evolution in the circuit model. Because this is the entire
population, the class balance and the class overlap are exact rather than estimated.
The main conclusion is conceptual rather than about any particular model. Which
decomposition is cheaper appears to be controlled primarily by a property of the
Hamiltonian's Pauli representation, its number of terms $n_{\mathrm{Pauli}}$, rather
than by conventional graph topology. This feature is the single most informative on its
own (MCC $0.423$), the degree variance is a useful complement to it, and the ten
topological features together carry almost none of the signal. Machine learning enters
as a practical way to learn this decision boundary. Across a range of models the
resulting MCC lands in a narrow band, from $0.569$ for the best untuned model up to
$0.593$ after tuning, with the top several models separated by less than the
seed-to-seed variation, so the finding is the feature dependence rather than any
particular architecture. It is also of practical value, since $n_{\mathrm{Pauli}}$ is
far cheaper to obtain than the CX count it predicts, by about a factor of $50{,}000$ at
$N=256$ ($0.003$\,s against $157$\,s per graph), so a compiler can consult it in place
of the expensive labeling it stands in for. For the downstream experiments we adopt one
model, a single-hidden-layer ANN of $32$ nodes at MCC $0.593$, and applied without
retraining
to the balanced held-out test set it reaches MCC $0.785$ at $N=8$ and rises to perfect
agreement for $N\ge64$, so a model tuned only on eight-vertex graphs chooses the
cheaper decomposition reliably on graphs an order of magnitude larger.

The feature-importance ranking of Section~\ref{sec:eda} does not depend on
the model used to produce it. We compute it with the tuned ANN, the model we ship,
and repeating the analysis with a gradient-boosting model, which is invariant to
feature scaling, yields the same two features as most important ($n_{\mathrm{Pauli}}$ at
$0.547$ and degree variance at $0.152$) and all others far smaller. The same gap
appears when each feature is examined on its own. The distributions of the
topological features are nearly identical for the two classes, so used individually
to rank graphs they achieve an ROC-AUC between $0.50$ and $0.54$, near random,
whereas $n_{\mathrm{Pauli}}$ reaches $0.94$ and $n_{\mathrm{match}}$ $0.58$. This is
why the matching-wins concentrate at low Pauli term counts while no single
topological feature separates the classes.

One might worry that the two decomposition counts are too close to the quantity we
predict, since they come from the same decompositions whose CX cost defines the
label. Two facts show that they do not trivially give away the answer. First, they
are much cheaper to compute than the label itself, as counting the terms of the two
decompositions costs about $O(N^{1.5})$ per graph, whereas measuring the true CX
counts requires building and transpiling both circuits at a cost of $O(N^{3.2})$
(Section~\ref{sec:features}). Second, the counts are not
equivalent to the label. Even the best model that uses them reaches only MCC
$0.593$, well short of perfect prediction, so they are useful but imperfect signals of
which decomposition is cheaper rather than a restatement of it.

There are some limitations that bound these conclusions and indicate the natural next steps.
The exhaustive population exists only at $N=8$, and the larger test graphs come from two datasets (from \cite{atallah2026simulating}), so generalization to CTQWs on other classes of graphs, and in particular families of graphs commonly used in CTQW search problems, remains untested. This limitation is sharpened by the composition of the transfer test noted in Section~\ref{sec:transfer}, where one family supplies almost all the Pauli-wins and the other almost all the matching-wins, so the strong transfer scores at large $N$ cannot yet be separated from a distinction between the two families. Establishing that the learned boundary is universal rather than family-specific requires evaluation on additional families that contain both classes. The size at which we
can validate our models is limited not by the features, but by the labeling cost, which grows as $O(N^{3.2})$ and reaches about four hours for a single graph at
$N=1024$, so we make no claim beyond $N=256$. We model cost as the CX gate count alone, leaving circuit depth, hardware connectivity, and swap insertion for a multi-objective classifier, and all labels come from a single reference implementation~\cite{atallah2026simulating}, so a study across transpiler optimization levels and additional graph families would establish how far the learned rule generalizes. The broader message is that whether the Pauli or the matching decomposition needs fewer CX gates can be predicted largely from inexpensive proxies, led by the number of terms in the Pauli decomposition and complemented by the degree variance, which suggests that other surrogates for CX gate count, or deriving the decision boundary in closed form,
is a promising research direction.

%==============================================================================

\section*{Author contributions}
M.A. designed and implemented the machine learning models and ran the experiments.
R.H. and Z.S. developed the idea to use machine learning to predict when CTQW simulation is efficient and secured funding. All authors wrote, read, edited, and approved the final manuscript.

\section*{Acknowledgments}
M. Atallah and R. Herrman acknowledge DE-SC0024290. Z. Saleem acknowledges DOE-145-SE-14055-CTQW-FY23. M. Atallah gratefully acknowledges the support of the Givens Associates program at Argonne National Laboratory. The funder played no role in study design, data collection, analysis and interpretation of data, or the writing of this manuscript.

\section*{Competing interests}
All authors declare no financial or non-financial competing interests. 

\section*{Reproducibility}\label{sec:repro}
Every number in Tables~\ref{tab:data} and \ref{tab:ladder}--\ref{tab:persize} is
measured, not illustrative. The permutation-importance analysis, the full-feature
model comparison, the per-model tuning sweep, the per-size transfer test, and
multi-seed intervals are each produced by a script over the released feature
matrix, and every reported value is read from the corresponding output file; where
a value is a mean over seeds, it is the mean over the same five fixed seeds
throughout. The timings reported in the main text, the per-graph feature and
labeling costs of Table~\ref{tab:cost}, the $O(N^{1.8})$ feature scaling and
$O(N^{3.2})$ labeling scaling of Sections~\ref{sec:dataset}
and~\ref{sec:features}, and the labeling-cost projection to larger $N$, were all
measured on a single core of an Intel Core i5-13420H
($8$ physical cores, $2.1$\,GHz base) with $40$\,GB of RAM, running Windows 11 and
Python $3.10$ with Qiskit $2.4$, scikit-learn $1.7$, and NumPy $2.2$. Code, the
labeled dataset, the trained model, the exact splits, and these result files are
available at \url{https://github.com/Mostafa-Atallah2020/ctqw-decomp-selector}.

%==============================================================================
\appendices

\section{Additional Metrics}\label{sec:appendix-metrics}
%==============================================================================
For completeness, Tables~\ref{tab:ladder-appendix}--\ref{tab:persize-appendix}
report the four metrics omitted from the corresponding main-text tables:
precision, recall, specificity, and negative predictive value (NPV). In terms of
the confusion-matrix counts of Section~\ref{sec:model} these are
\begin{align*}
\text{precision} &= \tfrac{T_p}{T_p+F_p}, &
\text{recall} &= \tfrac{T_p}{T_p+F_n}, \\
\text{specificity} &= \tfrac{T_n}{T_n+F_p}, &
\text{NPV} &= \tfrac{T_n}{T_n+F_n}.
\end{align*}
Row order and models match the corresponding main-text tables.

\begin{table*}[t]
\caption{Other metrics for the model comparison (Table~\ref{tab:ladder}).}
\label{tab:ladder-appendix}
\centering
\setlength{\tabcolsep}{6pt}
\begin{tabular}{lcccc}
\toprule
model & prec. & recall & spec. & NPV \\
\midrule
\textbf{gradient boosting} & \pms{0.766}{0.06} & \pms{0.445}{0.059} & \pms{0.994}{0.003} & \pms{0.978}{0.002} \\
logistic & \pms{0.742}{0.027} & \pms{0.436}{0.059} & \pms{0.994}{0.002} & \pms{0.978}{0.002} \\
LightGBM & \pms{0.645}{0.025} & \pms{0.436}{0.046} & \pms{0.99}{0.001} & \pms{0.978}{0.002} \\
random forest & \pms{0.715}{0.051} & \pms{0.393}{0.076} & \pms{0.993}{0.003} & \pms{0.976}{0.003} \\
ANN & \pms{0.645}{0.08} & \pms{0.433}{0.051} & \pms{0.99}{0.003} & \pms{0.978}{0.002} \\
XGBoost & \pms{0.609}{0.032} & \pms{0.45}{0.041} & \pms{0.988}{0.002} & \pms{0.978}{0.002} \\
decision tree ($d{=}3$) & \pms{0.695}{0.087} & \pms{0.362}{0.148} & \pms{0.992}{0.006} & \pms{0.975}{0.006} \\
SVM (RBF) & \pms{0.96}{0.034} & \pms{0.183}{0.027} & \pms{1}{0} & \pms{0.969}{0.001} \\
$k$-NN & \pms{0.726}{0.108} & \pms{0.212}{0.048} & \pms{0.997}{0.001} & \pms{0.97}{0.002} \\
decision tree ($d{=}2$) & \pms{0.821}{0.135} & \pms{0.188}{0.086} & \pms{0.997}{0.003} & \pms{0.969}{0.003} \\
naive Bayes & \pms{0.27}{0.043} & \pms{0.329}{0.036} & \pms{0.964}{0.006} & \pms{0.973}{0.001} \\
decision tree ($d{=}1$) & \pms{0.092}{0.184} & \pms{0.081}{0.162} & \pms{0.996}{0.008} & \pms{0.965}{0.006} \\
\bottomrule
\end{tabular}
\end{table*}

\begin{table*}[t]
\caption{Other metrics for the feature-subset ablation
(Table~\ref{tab:ablation}).}
\label{tab:ablation-appendix}
\centering
\setlength{\tabcolsep}{6pt}
\begin{tabular}{lcccc}
\toprule
feature subset & prec. & recall & spec. & NPV \\
\midrule
topological only ($10$) & \pms{0.467}{0.323} & \pms{0.01}{0.005} & \pms{1}{0} & \pms{0.962}{0} \\
$n_{\mathrm{Pauli}}$ only & \pms{0.659}{0.04} & \pms{0.293}{0.051} & \pms{0.994}{0.002} & \pms{0.972}{0.002} \\
$n_{\mathrm{Pauli}}+n_{\mathrm{match}}$ & \pms{0.651}{0.029} & \pms{0.312}{0.075} & \pms{0.993}{0.002} & \pms{0.973}{0.003} \\
$n_{\mathrm{Pauli}}+$ degree var. & \pms{0.771}{0.026} & \pms{0.448}{0.049} & \pms{0.995}{0.001} & \pms{0.978}{0.002} \\
\textbf{full set ($12$)} & \pms{0.774}{0.036} & \pms{0.476}{0.066} & \pms{0.994}{0.002} & \pms{0.979}{0.003} \\
\midrule
threshold on $n_{\mathrm{Pauli}}$ & \pms{0.47}{0.047} & \pms{0.512}{0.082} & \pms{0.976}{0.007} & \pms{0.981}{0.003} \\
\bottomrule
\end{tabular}
\end{table*}

\begin{table*}[t]
\caption{Other metrics for the per-model tuned sweep
(Table~\ref{tab:tuned}).}
\label{tab:tuned-appendix}
\centering
\setlength{\tabcolsep}{6pt}
\begin{tabular}{lcccc}
\toprule
model & prec. & recall & spec. & NPV \\
\midrule
\textbf{ANN} & \pms{0.774}{0.036} & \pms{0.476}{0.066} & \pms{0.994}{0.002} & \pms{0.979}{0.003} \\
gradient boosting & \pms{0.772}{0.049} & \pms{0.46}{0.034} & \pms{0.994}{0.002} & \pms{0.979}{0.001} \\
random forest & \pms{0.48}{0.024} & \pms{0.698}{0.051} & \pms{0.97}{0.003} & \pms{0.988}{0.002} \\
XGBoost & \pms{0.689}{0.038} & \pms{0.45}{0.03} & \pms{0.992}{0.002} & \pms{0.978}{0.001} \\
LightGBM & \pms{0.711}{0.032} & \pms{0.426}{0.048} & \pms{0.993}{0.002} & \pms{0.978}{0.002} \\
decision tree & \pms{0.332}{0.021} & \pms{0.638}{0.014} & \pms{0.949}{0.005} & \pms{0.985}{0.001} \\
SVM (RBF) & \pms{0.227}{0.01} & \pms{0.91}{0.027} & \pms{0.877}{0.004} & \pms{0.996}{0.001} \\
logistic & \pms{0.225}{0.012} & \pms{0.907}{0.047} & \pms{0.876}{0.004} & \pms{0.996}{0.002} \\
$k$-NN & \pms{0.629}{0.011} & \pms{0.279}{0.029} & \pms{0.993}{0.001} & \pms{0.972}{0.001} \\
naive Bayes & \pms{0.271}{0.044} & \pms{0.329}{0.036} & \pms{0.964}{0.006} & \pms{0.973}{0.001} \\
\bottomrule
\end{tabular}
\end{table*}

\begin{table*}[t]
\caption{Other metrics for the per-size transfer test
(Table~\ref{tab:persize}).}
\label{tab:persize-appendix}
\centering
\setlength{\tabcolsep}{5pt}
\begin{tabular}{rcccc}
\toprule
$n$ & prec. & recall & spec. & NPV \\
\midrule
8 & \pms{0.911}{0.059} & \pms{0.87}{0.068} & \pms{0.91}{0.066} & \pms{0.879}{0.054} \\
16 & \pms{0.807}{0.025} & \pms{0.924}{0.032} & \pms{0.779}{0.034} & \pms{0.912}{0.032} \\
32 & \pms{0.993}{0.005} & \pms{1}{0} & \pms{0.993}{0.005} & \pms{1}{0} \\
64 & \pms{1}{0} & \pms{1}{0} & \pms{1}{0} & \pms{1}{0} \\
128 & \pms{1}{0} & \pms{1}{0} & \pms{1}{0} & \pms{1}{0} \\
256 & \pms{1}{0} & \pms{1}{0} & \pms{1}{0} & \pms{1}{0} \\
\bottomrule
\end{tabular}
\end{table*}

%==============================================================================
\bibliographystyle{IEEEtran}
\bibliography{refs}

\end{document}